\documentclass[preprint,12pt]{elsarticle}
\usepackage{txfonts}
\usepackage{mathrsfs}
\usepackage{graphicx}
\usepackage{amssymb}
\usepackage{lineno}
\journal{Annals of Physics}
\begin{document}
\begin{frontmatter}
\title{Yangian symmetry in molecule \{V6\} and four-spin Heisenberg model}
\author{Xu-Biao Peng\corref{cor1}}
\ead{xubiaopeng@gmail.com}\cortext[cor1]{}
\author{Cheng-Ming Bai}
\author{Mo-Lin Ge\corref{cor1}}
\ead{geml@nankai.edu.cn}
\address{Theoretical Physics Division, Chern Institute of Mathematics ,Nankai University, Tianjin 300071,
P.R.China}
\begin{abstract}
The symmetry operator $Q=Y^2$ is introduced to re-describe the
Heisenberg spin triangles in the \{V6\} molecule, where $\mathbf{Y}$ stands for the Yangian operator which can be viewed as special form of  Dzyaloshiky-Moriya (DM) interaction for spin 1/2 systems. Suppose a parallelogram Heisenberg model that is comprised of four $\frac{1}{2}$-spins commutes with $Q$, which means that it possesses Yangian symmetry, we show that the ground state of the Hamiltonian $H_4$ for the model allows to take the total spin $S=1$ by choosing some suitable exchange constants in $H_4$. In
analogy to the molecular \{V6\} where the two triangles interact through Yangian operator we then give the magnetization for the theoretical molecule ``\{V8\}'' model which is comprised of two parallelograms. Following the example of molecule \{V15\}, we give another theoretical molecule model regarding the four $\frac{1}{2}$-spins system with total spin $S=1$ and predict the local moments to be
$\frac{9}{10}\mu_{B}$, $\frac{1}{10}\mu_{B}$, $\frac{1}{10}\mu_{B}$,
$\frac{9}{10}\mu_{B}$ respectively.
\end{abstract}


\end{frontmatter}

\section{Introduction}\label{sec-1}
The single molecular magnets(SMMs)have attracted much attention both
for its scientific importance of studying fundamental issues and for
its potential applications. The magnetic molecules \{V6\} and
\{V15\} provide us a good platform for exploring these issues for the models whose total spins are not large. There have been beautiful
investigations on these respects.\cite{c,b,a,d} \\

\begin{figure}
\centering
\includegraphics[width=5cm,height=5cm]{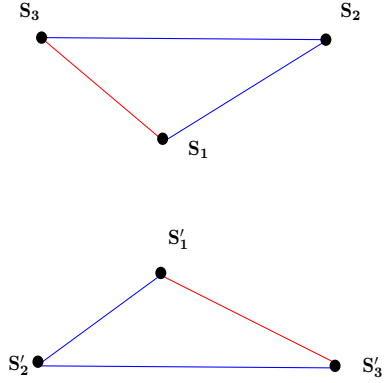}
\caption{(color online)The structure model of \{V6\} molecule. The
blue lines represent the two equal exchange interaction constants
called $J_{a}$, and the red line is the third exchange interaction
constant $J_{c}$. There is DM interaction $H_{inter}=\Delta(\mathbf
S_{A}\times \mathbf{S}_{A^{\prime}})_{y}$ between the two
triangles.\label{V6}}
\end{figure}

For latter use, let us briefly introduce the structures of the
molecule \{V6\} first. As was mentioned in the Ref.\cite{b},
the molecular \{V6\} is the abbreviation of the molecule
(CN$_3$H$_6$)$_4$Na$_2$[H$_{4}$V$_{6}$O$_{8}$(PO$_{4}$)$_{4}$\{(OCH$_{2}$)$_{3}$CCH$_{2}$OH\}$_{2}$]$\cdot$
14H$_2$O whose structure is shown in Fig.\ref{V6}. We see that the
molecule consists of two pieces, each of which is an isosceles
triangle. In each triangle, two of the spin exchange constants are
equal(shown in blue $J_{a}\sim65K$) and the third one is smaller(shown in red $J_{c}\sim7K$).\cite{a} The experiment
has shown that there is a kind of special interaction called
Dzyaloshiky-Moriya (DM) interaction\cite{DM} between the two triangles,
whose Hamiltonian can be written as $H_{inter}=\Delta(\mathbf
S_{A}\times \mathbf{S}_{A^{\prime}})_{y}$.\cite{b} The operators
$\mathbf S_{A}$ and $\mathbf S_{A^{\prime}}$ are the total spin
operators of two triangles, respectively, and the energy gap
$\Delta$ is tiny. However, such an interaction can make
contributions to the Landau-Zener-St\"uckelberg (LZS)
transition\cite{j} when the magnetic field is absent. The LZS effect can be detected by measuring the magnetization of the molecule
\{V6\}.\cite{b}\\

In Ref.\cite{c,b,a,d}, the wave functions of molecule \{V6\}
and the experimental measurements of the system have been shown
clearly. However, from the theoretical point of view, each triangle
of the molecule \{V6\} is formed by three spins, and the symmetry
properties of the triangle desire to be investigated. In this
article, we first focus on triangle of a molecule \{V6\}, and
introduce a new symmetry operator $Q$ to re-describe the triangular
piece. We can see that the commutativity between such a new symmetry
operator $Q$ and the Hamiltonian of the triangle $H_3$ will constrain the parameters
in $H_3$ leading to $J_{12}=J_{23}$(see Eq.(\ref{relationtriangle})). The symmetry operator $Q$ looks a natural description
for the triangle model in the molecular \{V6\}. Further we shall
extend such a new symmetry to a four-spin Heisenberg model. Using the extended
symmetry operator $Q$ in the four-spin system, the Hamiltonian $H_4$ for a parallelogram will be restricted. By
analyzing the Hamiltonian $H_4$, we find that the ground state can
be with total spin $S=1$ in some special cases, which has not been
considered before. Based on this assumption, we make a
prediction on the magnetization of a theoretical molecule ``\{V8\}'' which is comprised of two parallelograms with the DM interaction between them
in analogy to the molecule \{V6\}.\cite{b} To test the
parallelogram model itself, we propose another molecular model in analogy to \{V15\} and give the prediction of its local spin moments configuration.\\

This article is organized as follows: In Sec.\ref{sec-2},we shall
investigate the single Heisenberg spin triangle model in \{V6\} and
introduce the new quantum number operator $Q$ to re-describe the model. We
shall show that the operator $Q$ represents a new symmetry in the
three-spin system. In Sec.\ref{sec-3}, we shall extend the symmetry
to a system comprised of four $\frac{1}{2}$-spins and
establish its Hamiltonian $H_4$. We demonstrate that the ground
state of $H_4$ allows to be with total spin $S=1$ in some special cases.
Besides, we make up a theoretical molecular model called ``\{V8\}'' in
which the Yangian interaction between the two parallelograms is introduced and
the prediction of its magnetization is made. In
Sec.\ref{sec-4}, we discuss another molecular model which contains
only one parallelogram and give the prediction about its local spin
moments configuration. In the Appendix, we shall show details about
how the symmetry operator determines the Hamiltonian.

\section{The new symmetry in \{V6\}\label{sec-2}}

As was mentioned in the Sec.\ref{sec-1}, a molecule \{V6\} is
comprised of two triangles with a tiny DM interaction between them.
In this section, we shall concentrate on only one triangular piece in
a molecule \{V6\}. It has been verified that the triangle model
is actually isosceles by the experiment.\cite{a} Hence the \{V6\} problem has been well-established both theoretically and experimentally. However, in this section
we would like to introduce a new symmetry operator $Q$ to re-describe
the triangle system that can be extended to more $\frac{1}{2}$-spins system.\\

The model of the Heisenberg spin triangle is comprised of there spin-$\frac{1}{2}$ particles, whose Hamiltonian is written
as:
\begin{eqnarray}\label{tri}
H_{0}=J_{12}\mathbf S_{1}\cdot \mathbf S_{2}+J_{23}\mathbf
S_{2}\cdot \mathbf S_{3}+J_{13}\mathbf S_{1}\cdot \mathbf S_{3},
\end{eqnarray}
where $\mathbf S_{1}$, $\mathbf S_{2}$ and $\mathbf S_{3}$ are the
spin operators of three particles, and the relationship for the
interaction constants $J_{12}$, $J_{13}$ and $J_{23}$ is unknown. If
a magnetic field is applied along $z$ axis, then the term
corresponding to the Zeeman Energy $H_{Zeeman}=\mu
B(S_{1z}+S_{2z}+S_{3z})$ should be included in Eq.(\ref{tri}).The
Zeeman term can split the energy levels with different eigenvalues of
$S_{z}$, where $\mathbf S=\mathbf S_{1}+\mathbf S_{2}+\mathbf S_{3}$
is the total spin operator of the triangle. Obviously, only the
quantum numbers $S^{2}$ and $S_{z}$ are not adequate to describe a
system with three spins. It is easy to see that there are two different
eigenstates corresponding to the same quantum numbers $S=1/2,S_{z}=-1/2$.\\

The new symmetry operator $Q$ that we shall introduce is written as
$Q=Y^2$, where the operator $\mathbf{Y}$ is a special form of the DM
interaction in the Heisenberg spin triangle written as:
\begin{eqnarray}\label{Yangian}
\mathbf Y=i(\mathbf S_{1}\times\mathbf S_{2}+\mathbf
S_{2}\times\mathbf S_{3}+\mathbf S_{1}\times\mathbf S_{3}).
\end{eqnarray}
It can be verified that such a new operator satisfies the
commutation rules as $\lbrack Q,S^{2}\rbrack=0$ and $\lbrack
Q,S_{z}\rbrack=0$. So we shall use the operator $Q$ to represent
certain symmetry property of the three-spin system just like what we
have done in the Hydrogen Atom.\cite{HA} (In Mathematics Physics,
the operator $\mathbf{Y}$ is in fact a special form of the Yangian
operator, see Appendix.\ref{mathYang}.) The operator $Q$ can be
viewed as a collective quantum number that describes the history
besides $S^2$ and $S_{z}$. If we take the set $\{S^2,S_{z},Q\}$ to
be the complete operator set of the system, the commutativity
$\lbrack Q,H\rbrack=0$ is wanted to be satisfied. Based on such a constrain, we
can easily get
\begin{eqnarray}\label{relationtriangle}
J_{12}=J_{23}.
\end{eqnarray}
Fortunately, this relationship
$J_{12}=J_{23}$ has been shown to exist in molecule
\{V6\}\cite{demon1}. With this symmetry the Hamiltonian in
Eq.(\ref{tri}) can be simplified as:
\begin{eqnarray}
H_{3}=J_{12}(\mathbf S_{1}\cdot \mathbf S_{2}+\mathbf S_{2}\cdot
\mathbf S_{3})+J_{13}\mathbf S_{1}\cdot \mathbf S_{3},
\end{eqnarray}
We emphasize that in the triangle model, the eigenvectors of the
operator $Q$ are nondegenerate, so the Yangian symmetry operator
$Q$ can uniquely determine the Hamiltonian of the system. The complete set $\{S^{2},S_{z},Q\}$ can be used to determine the states in
triangular piece in the \{V6\} model described in
Sec.\ref{sec-1}. By directly diagonalizing the matrix $Q$ in the
usual Lie Algebraic representation (see Appendix.\ref{thspin}), we
get the two states with total spin $S=\frac{1}{2},S_{z}=-\frac{1}{2}$ as:
\begin{eqnarray}
\vert\phi_{\alpha}\rangle&=&\frac{-1}{\sqrt{6}}\left(\vert\uparrow\downarrow\downarrow\rangle+\vert\downarrow\downarrow\uparrow\rangle-2\vert\downarrow\uparrow\downarrow\rangle\right)\label{states1},\\
\vert\phi_{\beta}\rangle&=&\frac{1}{\sqrt{2}}\left(\vert\uparrow\downarrow\downarrow\rangle-\vert\downarrow\downarrow\uparrow\rangle\right)\label{states2},
\end{eqnarray}
which are corresponding to the eigenvalues
$-\frac{1}{4},-\frac{9}{4}$ of $Q$ respectively. It can be verified
that these states are the eigenstates of the Hamiltonian $H_{3}$,
and the corresponding levels are
$E_{\frac{1}{2},-\frac{1}{2}}^{\alpha}=\frac{1}{4}J_{13}-J_{12}$ and
$E_{\frac{1}{2},-\frac{1}{2}}^{\beta}=-\frac{3}{4}J_{13}$.\\

Hence, the special DM interaction operator $\mathbf{Y}$ plays an
important role in the Heisenberg spin triangle in \{V6\}. Its square
$Q$ is a new quantum number operator representing the Yangian symmetry. Using $Q$ we can easily determine
the form of the Hamiltonian and directly obtain the eigenstates
of the system. Therefore it is reasonable to extend such a Yangian symmetry to
a four-spin system that will be discussed in the next section.\\

\section{Four-spin Heisenberg model}\label{sec-3}

\subsection{A four-spin system determined by the Yangian symmetry}

\begin{figure} \centering
\includegraphics[width=4cm]{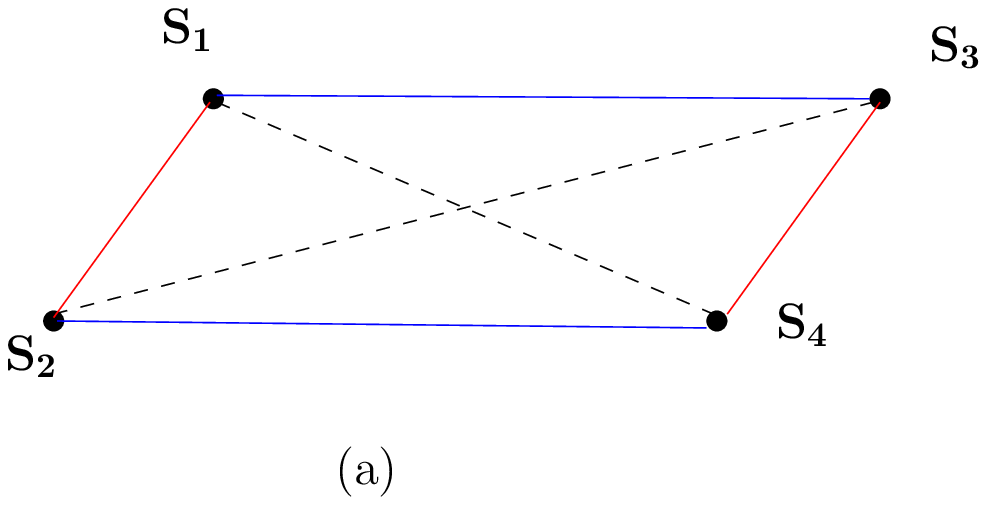}
\includegraphics[width=4.5cm]{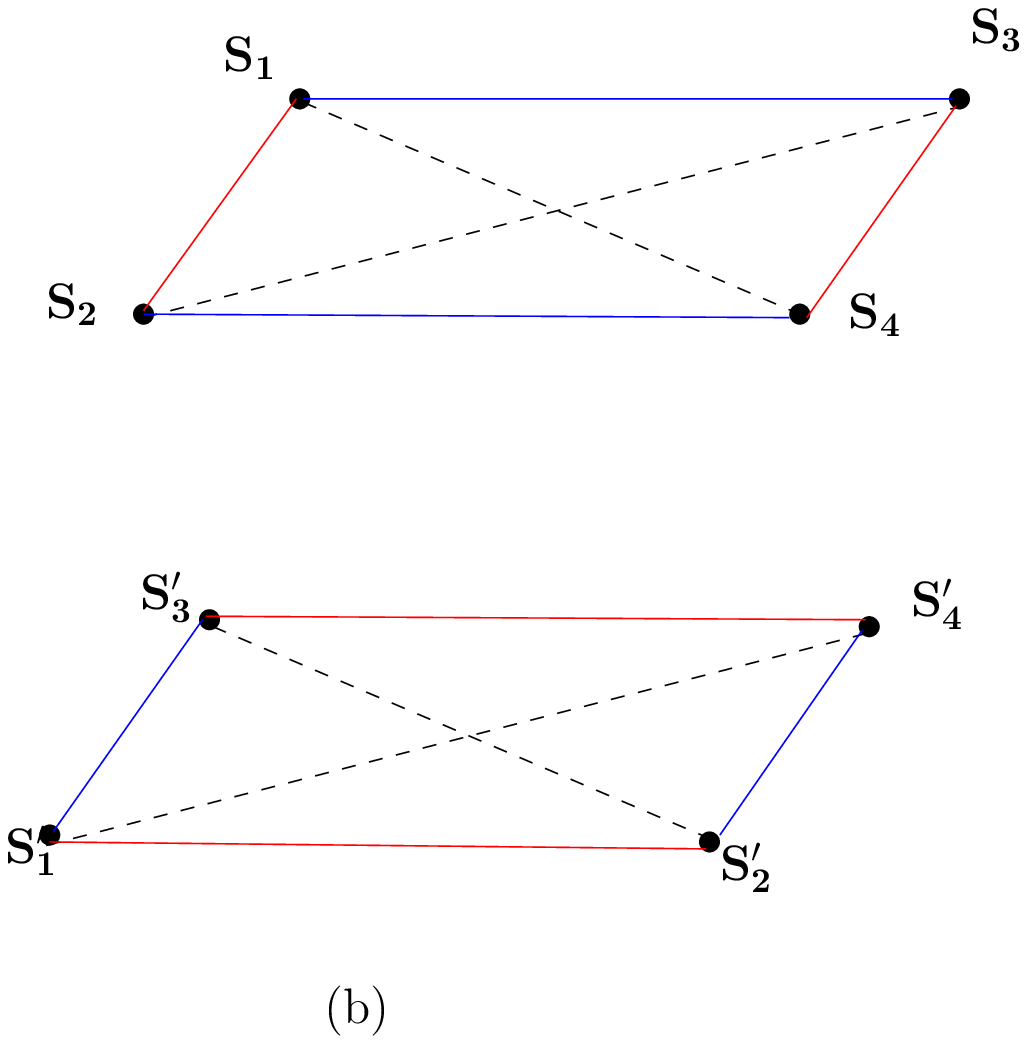}
\caption{(a) (color online) The model of the four-spin system with
the Yangian symmetry, the red line represents the exchange constant
$a_{12}$, and the blue represent $a_{13}$. (b) (color online) The
model of the extended \{V6\}-type molecule, the interaction between
the two parallelograms is $H_{inter}=\Delta(\mathbf S_{A}\times
\mathbf{S}_{A^{\prime}})_{y}$, which is totally similar to the
\{V6\} molecule. \label{fig2}}
\end{figure}

For a system comprised of four particles with each spin $\frac{1}{2}$ the
general form of the Hamiltonian can be written as:
\begin{eqnarray}\label{4Hamilton}
H=\sum^4_{j>i=1}a_{ij}\mathbf S_{i}\cdot\mathbf S_{j},
\end{eqnarray}
and the Yangian operator is
defined by:
\begin{eqnarray}\label{4Yangian}
\mathbf Y=i\sum^4_{j>i=1}(\mathbf S_{i}\times\mathbf S_{j}).
\end{eqnarray}
which is identical with the special DM interaction operator for the four-spin system. For a four-spin system with the Yangian symmetry $Q=Y^{2}$ should be a quantum number operator. Similar to the
three-spin system in the Sec.\ref{sec-2}, we shall get the
constrain for the parameters in Eq.(\ref{4Hamilton}) based on the
commutativity $\lbrack Q,H\rbrack=0$. To calculate the commutation
relation, we need get the eigenstates of $Q$ first, and then let
the Hamiltonian $H$ share the same eigenstates with $Q$. The eigenvalues and eigenvectors of $Q$ are shown as Eq.(\ref{eigeneqation}) in the
Appendix.\ref{fspin}. It should be noted that the eigenvectors $|\psi_{1,m}^1\rangle$ and $|\psi_{1,m}^3\rangle$ are degenerate in Eq.(\ref{eigeneqation}). The eigenstates with eigenvalue $Q=-\frac{1}{2}$ can be linear combinations of the two degenerate states. In fact, the combination is a $SU(2)$ rotation on the eigenvectors $|\psi_{1,m}^1\rangle$ and $|\psi_{1,m}^3\rangle$ because of the orthogonality and normalization of the eigenvectors. So we need to introduce an additional variable $\theta$ to indicate the general eigenstates with $Q=-\frac{1}{2}$ as follows:
\begin{eqnarray}\label{degenerate}
\left(\begin{array}{c}|\psi_{1,m}^{1^\prime}\rangle\\|\psi_{1,m}^{3^\prime}\end{array}\right)=\left(\begin{array}{cc}\cos\frac{\theta}{2}& -\sin\frac{\theta}{2}\\ \sin\frac{\theta}{2}& \cos\frac{\theta}{2}\end{array}\right)\left(\begin{array}{c}|\psi_{1,m}^{1}\rangle\\|\psi_{1,m}^{3}\end{array}\right)
\end{eqnarray}
In this situation, only the eigenstates as Eq. (\ref{degenerate}) can be considered as the ones of the Hamiltonian $H$ with quantum number $Q=-\frac{1}{2}.$ Letting $H$ share the same eigenvectors with $Q$, through careful calculation shown in Appendix.\ref{fspin}, we obtain the Hamiltonian of the four-spin system as Eq. (C.10). In a special case ($\theta=0$), we find
the relations for the exchange interaction constants in
Eq.(\ref{4Hamilton}) due to the $Q$-symmetry:
\begin{eqnarray}\label{relation}
a_{12}&=&a_{34},\qquad a_{13}=a_{24},\nonumber\\
a_{14}&=&\frac{1}{3}\left(a_{12}+2a_{13}\right),\nonumber\\
a_{23}&=&\frac{5}{3}a_{12}-\frac{2}{3}a_{13}.
\end{eqnarray}
So the Hamiltonian can be written as a simpler form
\begin{eqnarray}\label{4HM}
H_4&=&a_{12}(\mathbf S_{1}\cdot\mathbf S_{2}+\mathbf
S_{3}\cdot\mathbf S_{4})+a_{13}(\mathbf S_{1}\cdot\mathbf
S_{3}+\mathbf
S_{2}\cdot\mathbf S_{4}){}\nonumber\\
&&{}+\frac{1}{3}(a_{12}+2a_{13})\mathbf S_{1}\cdot\mathbf
S_{4}+\frac{1}{3}(5a_{12}-2a_{13})\mathbf S_{2}\cdot\mathbf S_{3}.\nonumber\\
&&
\end{eqnarray}
The Hamiltonian shown by Eq. (\ref{4HM}) is actually a parallelogram
model\footnote{In fact, the model of such a system is generally a
tetrahedron, but here we only take the simplest case of a
parallelogram.} as shown in Fig.\ref{fig2}(a), where there are
only two independent exchange constants called $a_{12}$ and $a_{13}$.

The eigenvalues and eigenstates of the Hamiltonian Eq. ({\ref{4HM}}) are as follows:
\begin{eqnarray}
H_4|\psi_{2,m}\rangle&=&E_{2,m}|\psi_{2,m}\rangle=(a_{12}+\frac{1}{2}a_{13})|\psi_{2,m}\rangle,\nonumber\\
H_4|\psi^{1}_{1,m}\rangle&=&E_{1,m}^1|\psi^{1}_{1,m}\rangle=-\frac{1}{2}a_{13}|\psi^{1}_{1,m}\rangle,\nonumber\\
H_4|\psi^{2}_{1,m}\rangle&=&E_{1,m}^2|\psi^{2}_{1,m}\rangle=(\frac{1}{3}a_{12}-\frac{5}{6}a_{13})|\psi^{2}_{1,m}\rangle,\nonumber\\
H_4|\psi^{3}_{1,m}\rangle&=&E_{1,m}^3|\psi^{3}_{1,m}\rangle=(-\frac{4}{3}a_{12}+\frac{5}{6}a_{13})|\psi^{3}_{1,m}\rangle,\nonumber\\
H_4|\psi^{+}_{0,0}\rangle&=&E_{0,0}^+|\psi^{+}_{0,0}\rangle=(-2a_{12}+\frac{1}{2}a_{13})|\psi^{+}_{0,0}\rangle,\nonumber\\
H_4|\psi^{-}_{0,0}\rangle&=&E_{0,0}^-|\psi^{-}_{0,0}\rangle=-\frac{3}{2}a_{13}|\psi^{-}_{0,0}\rangle,
\end{eqnarray}
where $m$ is the quantum number of $S_{z}$ and the eigenstates are
shown by Eq.(\ref{eigenvec}) in the Appendix.\ref{fspin}. If a
magnetic field is applied, the states with different quantum numbers
$m$ will split because of the Zeeman Effect. When the field is weak
enough, we emphasize that the singlet state is not
always the ground state. In fact, the ground state depends on the
values of the exchange constants. By choosing some suitable
constants, we can get the ground states with total spin $S=1$. For
example, when $a_{12}>0,a_{13}<0$, and $a_{13}<-2a_{12}$, the energy levels are $E^3_{1,m}<E^+_{0,0}<E_{2,m}<E^-_{0,0}<E^1_{1,m}<E^2_{1,m}$, so the ground state is
\begin{eqnarray*}
|\psi^{3}_{1,-1}\rangle=\frac{1}{\sqrt{20}}\left(|\uparrow\downarrow\downarrow\downarrow\rangle-3|\downarrow\uparrow\downarrow\downarrow\rangle+3
|\downarrow\downarrow\uparrow\downarrow\rangle-|\downarrow\downarrow\downarrow\uparrow\rangle\right).
\end{eqnarray*}
\\

Thus, we have got the Hamiltonian of the four-spin system determined
by the Yangian symmetry. In a special case it is a model of
parallelogram. It allows the
ground state with $S=1$ instead of $S=0$. This is an
interesting case not considered before, and we shall predict the ``\{V8\}''-type molecule based on this result and make more theoretical predictions.\\

\subsection{The magnetization of ``\{V8\}''-type molecule}

In the molecule \{V6\} we have shown that there is Yangian
symmetry characterized by the operator $Q$ for each triangle.
Meanwhile it has also been proved by the experiment\cite{b} that
there is a DM interaction between the two triangular pieces in
\{V6\}. It causes the LZS transition.\cite{j} The effect of the
LZS transition can be reflected through the magnetization of the
molecular \{V6\}.\\

\begin{figure} \centering
\includegraphics[width=7cm]{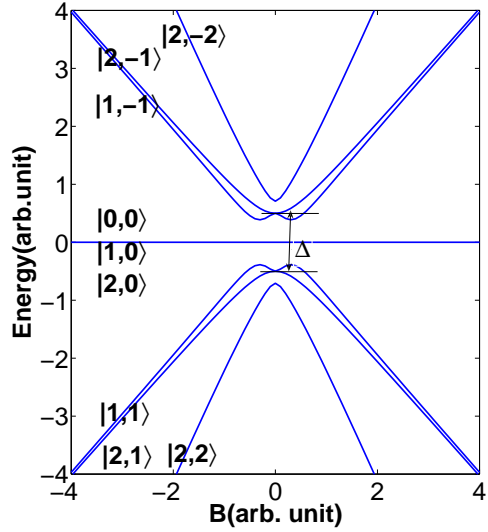} \caption{The scheme of the
energy levels of the system comprised of two parallelograms with a
DM interaction between them, where $\Delta$ is the energy gap of the
Landau-Zener tunneling.\label{fig3}}
\end{figure}

It is natural to extend the LZS transition for \{V6\} to a molecular
system comprised of two parallelograms. The molecular model is shown
in Fig.\ref{fig2}(b), which is called ``\{V8\}''. We also assume
that there is a similar DM interaction written as
$H_{inter}=\Delta(\mathbf S_{A}\times \mathbf{S}_{A^{\prime}})_{y}$
between the two parallelograms. The operators $\mathbf S_{A}$ and
$\mathbf S_{A^{\prime}}$ are the total spin operators of the two
parallelograms, respectively. Obviously, it is complicated to
calculate such a system with eight $\frac{1}{2}$-spins. To simplify the situation, we only consider the ground state of each parallelogram as in the
molecule \{V6\}.\cite{a} If the total spin of ground state is zero,
it will make no sense to study the magnetization. However, if the
total spin is one, we can make a prediction about its magnetization.
Extending the approach working well for \{V6\} model the Hamiltonian of ``\{V8\}'' model will be written as:
\begin{eqnarray}\label{Hm}
\mathscr{H}=\hbar\gamma B(t)(S_{Az}+S_{A^{\prime}z})+\Delta(\mathbf
S_{A}\times \mathbf{S}_{A^{\prime}})_{y},
\end{eqnarray}
where the spins $S_{A}=1$ and $S_{A^{\prime}}=1$, and $\gamma$ is
the gyromagnetic ratio. The magnetic field $B(t)$ is along the $z$
axis and varies with the time. The first term in Eq.(\ref{Hm}) is
about the Zeeman Energy, and the second one is the special DM
interaction leading to the LZS transition and is assumed to be tiny.
By exactly diagonalizing the Hamiltonian $\mathscr{H}$, we can get
the energy levels shown in Fig.\ref{fig3}.
The DM interaction plays an important role only near the ``crossing
point'' of the energy levels where the magnetic field is absent. Far
away from the ``crossing point'', the system behaves as two
independent particles with each spin $S=1$, so we just need to
consider one particle to investigate the magnetization of the
system. In the experiment, the measurement of the magnetization is
related to the relaxation time of the magnetization, and here we
assume the relaxation time is long compared with the experimental
time so that the hysteresis effect can be observed. The general
method to investigate the magnetization has been well discussed in
the Ref.\cite{c}, and we shall directly refer to the
conclusion in the Ref.\cite{c}. Using the standard formula (33) in Ref.\cite{c}
\begin{eqnarray*}
\frac{d\rho_{NN}}{dt}=\sum_{N^{\prime}}W_{N^{\prime}N}\left(t\right)\rho_{N^{\prime}N^{\prime}}-\sum_{N^{\prime}}W_{NN^{\prime}}\left(t\right)\rho_{NN}
\end{eqnarray*}
 where the meaning of $\rho$ and $W$ will be seen clearly later, we obtain the equation of the magnetization dynamic for spin $S=1$
\begin{eqnarray}\label{hysteresis}
\frac{d}{dt}\left(\begin{array}{c}\frac{M}{-\hbar\gamma}\\\rho_{00}\end{array}\right)=\left(\begin{array}{cc}C_{1}&C_{2}\\C_{3}&C_{4}\end{array}\right)
\left(\begin{array}{c}\frac{M}{-\hbar\gamma}\\\rho_{00}\end{array}\right)+\left(\begin{array}{c}E\\F\end{array}\right),
\end{eqnarray}
where $M$ is the magnetization of the system, and $\rho$ is the
density operator. The magnetization is written
as: $M=-\hbar\gamma\langle
S_{z}\rangle=-\hbar\gamma(\rho_{++}-\rho_{--})$ and the
coefficients in Eq.(\ref{hysteresis}) are as follows:
\begin{eqnarray*}
C_{1}&=&-\frac{1}{2}(W_{-0}-W_{+0})-W_{-+}-W_{+-},\\C_{2}&=&\frac{1}{2}(W_{+0}-W_{-0})+W_{0+}-W_{0-}+W_{+-}-W_{-+},\\
C_{3}&=&\frac{1}{2}(W_{+0}-W_{-0}),\\C_{4}&=&-\frac{1}{2}(W_{+0}+W_{-0}+2W_{0+}+2W_{0-}),\\
E&=&\frac{1}{2}(W_{-0}-W_{+0})+W_{-+}-W_{+-},\\F&=&\frac{1}{2}(W_{+0}+W_{-0}).
\end{eqnarray*}

\begin{figure} \centering
\includegraphics[width=6.5cm]{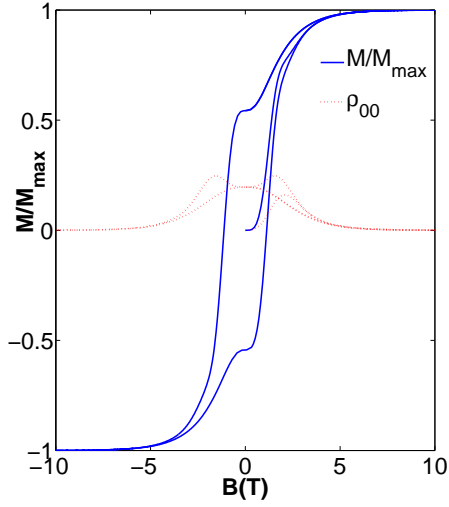}
\caption{(color online) The magnetization of the four-spin system in
a low temperature.The blue solid line is the magnetization, and the
red dashed line is the density matrix $\rho_{00}$. The function of
the magnetic field varies with the time as $B(t)=10\sin(t)$.
\label{fig4}}
\end{figure}

The quantity $W_{NN^{\prime}}$ represents the transition probability from the state $|N\rangle$ to $|N^\prime\rangle$, where $N$, $N^\prime$ is the magnetic quantum number characterized as $+,0,-$. If we take one-phonon process approximation the transition probability can be replaced by\cite{e}
\begin{eqnarray*}
W_{NN^{\prime}}=\frac{A\delta^{3}}{1-e^{-\beta\hbar\delta}},
\end{eqnarray*}
where $\delta=E_{N}-E_{N^{\prime}}=\hbar\gamma B(t)(N-N^{\prime})$
is the energy difference between two levels. When a pulsed
magnetic field is applied, a hysteresis loop(see
Fig.\ref{fig4}) can be obtained from the Eq.(\ref{hysteresis}),
where the magnetization has been normalized with
$M_{max}=\hbar\gamma$. All the parameters in the
Eq.(\ref{hysteresis}) have been taken the data provided in the Ref.\cite{b} for \{V6\} molecule. We see that it is the usual
hysteresis loop of a molecular magnet with the spin $S=1$.\\

However, the LZS transition must be considered when the magnetic field varies to the vicinity of the ``crossing point'' in the ``\{V8\}'' model. The total Hamiltonian is written as Eq.(\ref{Hm}), and the
off-diagonal elements can no longer be ignored. Diagonalizing Eq. (\ref{Hm}) the exact energy levels read
\begin{eqnarray*}
E_{2,0}&=&E_{1,0}=E_{0,0}=0;\\E_{2,\pm1}&=&\pm\sqrt{B^{2}(t)+\Delta^{2}};\\
E_{2,\pm2}&=&\pm\frac{1}{\sqrt{2}}\Bigg [5\mu^{2}B^{2}(t)+3\Delta^{2}+\sqrt{9\mu^{4}B^{4}(t)+30\mu^{2}B^{2}(t)+\Delta^{4}}\Bigg]^\frac{1}{2};\\
E_{1,\pm1}&=&\pm\frac{1}{\sqrt{2}}\Bigg[5\mu^{2}B^{2}(t)+3\Delta^{2}-\sqrt{9\mu^{4}B^{4}(t)+30\mu^{2}B^{2}(t)+\Delta^{4}}\Bigg]^\frac{1}{2},
\end{eqnarray*}
where the subscript $(l,m)$ corresponds to the state
$|l,m\rangle$ which is the asymptotic eigenstates when the field is strong.
Here $l$ is the quantum number of total spin, and $m$ is the
magnetic quantum number. For simplicity, we just choose the nearest
three levels $E_{2,-1},E_{2,0},E_{2,1}$ for a weak field to
investigate the effect of the off-diagonal elements. These three
levels can be viewed as the eigenvalues of the Hamiltonian with the
matrix form
\begin{eqnarray}
\mathscr{H}=\left(\begin{array}{ccc}B\left(t\right)&\frac{\Delta}{\sqrt{2}}&0\\\frac{\Delta}
{\sqrt{2}}&0&\frac{\Delta}{\sqrt{2}}\\0&\frac{\Delta}{\sqrt{2}}&-B\left(t\right)\end{array}\right),
\end{eqnarray}
whose eigenstates are
\begin{eqnarray}
|E_{+},t\rangle&=&\left(\frac{1+\cos\beta}{2},\frac{\sin\beta}{\sqrt{2}},\frac{1-\cos\beta}{2}\right)^{T},\nonumber\\
|E_{0},t\rangle&=&\left(\frac{-\sin\beta}{\sqrt{2}},\cos\beta,\frac{\sin\beta}{\sqrt{2}},\right)^{T},\nonumber\\
|E_{-},t\rangle&=&\left(\frac{1-\cos\beta}{2},-\frac{\sin\beta}{\sqrt{2}},\frac{1+\cos\beta}{2},\right)^{T},
\end{eqnarray}
where the angle $\beta$ is determined by
$\cos\beta=\frac{B\left(t\right)}{\sqrt{B^{2}\left(t\right)+\Delta^{2}}}$
and the corresponding eigenvalues are $E_{0}=0,\qquad
E_{\pm}=\pm\sqrt{B^{2}\left(t\right)+\Delta^{2}}$. Following the
method shown in Ref.\cite{c} and assuming the Landau-Zener
tunneling in adiabatic approximation, we derive the magnetization for spin $S=1$:
\begin{eqnarray}\label{Magnetic}
M=-\hbar\gamma\langle
S_{z}\rangle=-\hbar\gamma\left(\tilde{\rho}_{++}S^{++}_{z}+\tilde{\rho}_{00}S^{00}_{z}+\tilde{\rho}_{--}S^{--}_{z}\right).
\end{eqnarray}

\begin{figure} \centering
\includegraphics[width=6.5cm]{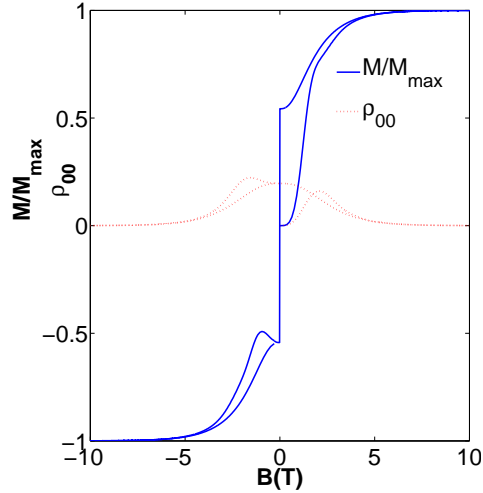}
\caption{(color online) The magnetization of the system ``\{V8\}'' comprised of
two parallelograms with DM interaction between them. The blue solid
line is the magnetization, and the red dashed line is the density
matrix $\rho_{00}$. It turns out that the density matrix $\rho_{00}$ is not
affected by the off-diagonal matrix of the Hamiltonian, and its
magnetization has a similar behavior to the molecular
\{V6\}.\cite{k}\label{hysteresis2}}
\end{figure}

It is easy to calculate the matrix elements of $S_{z}$ ,and the
Eq.(\ref{Magnetic}) can be simplified as
$M=\hbar\gamma\cos\beta\cdot n$. Here $n$ is defined as
$n=\tilde{\rho}_{--}-\tilde{\rho}_{++}$, and obeys the same Bloch
equation as Eq. (\ref{hysteresis}):
\begin{eqnarray}\label{den}
\frac{d}{dt}\left(\begin{array}{c}n\left(t\right)\\\tilde{\rho}_{00}\end{array}\right)=\left(\begin{array}{cc}C_{1}&C_{2}\\C_{3}&C_{4}\end{array}\right)
\left(\begin{array}{c}n\left(t\right)\\\tilde{\rho}_{00}\end{array}\right)+\left(\begin{array}{c}E\\F\end{array}\right).
\end{eqnarray}
Combining the Eq.(\ref{Magnetic}) with Eq. (\ref{den}), we can get the
magnetization of the system as shown in Fig. \ref{hysteresis2}, where
the applied magnetic field is taken as $B(t)=10\sin(t)$ varying in a
period $t\in [0,\pi]$. Similar to the \{V6\} molecule the
magnetization can be probed because of the LZS effect.\\

In a summary, we extended the properties of the molecule
\{V6\} to a ``\{V8\}''-type molecule comprised of two parallelograms. By
investigating its magnetization of the molecular model, we draw the
conclusion that the ``\{V8\}''-type molecule behaves similar to
the molecule \{V6\}.\\

\section{The local spin moments configuration of the four-spin system}\label{sec-4}

In this section, we will make up another molecule whose model can be
treated as one parallelogram and give a prediction about its local
spin moments configuration. \\

\begin{figure}
\centering
\includegraphics[width=3cm,height=5.5cm]{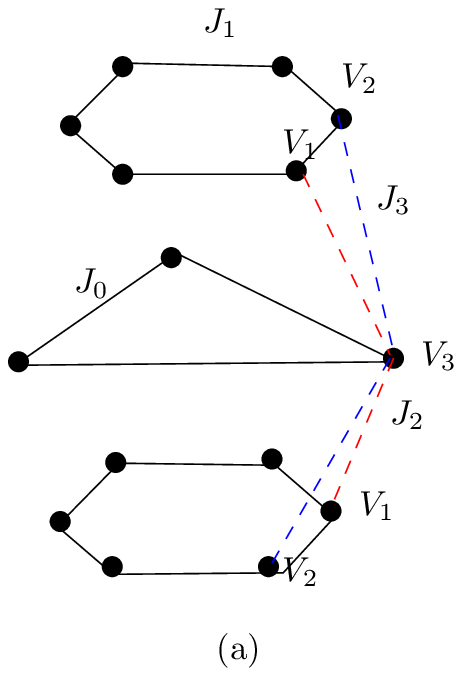}
\includegraphics[width=5.5cm]{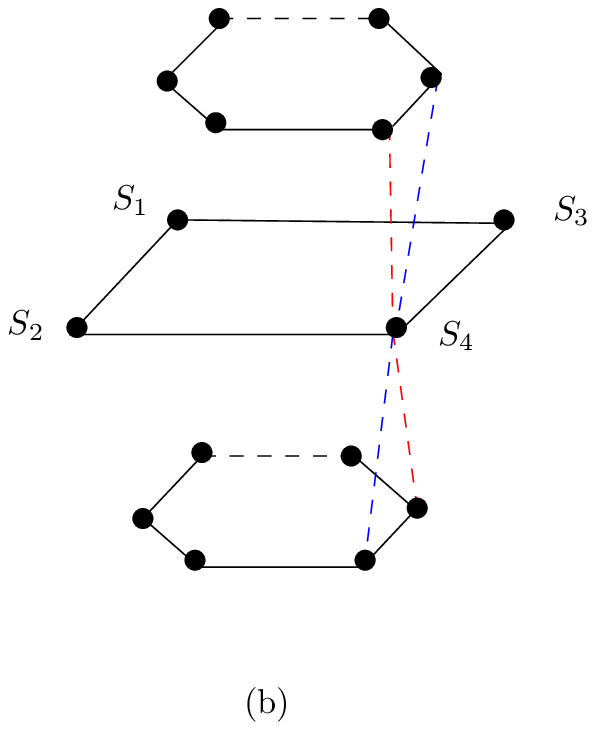}
\caption{(a)(color online) The structure model of the molecular
\{V15\}. The solid circles represent the $V^{4+}$($s=1/2$) ions. In
a low temperature, only the triangle in the middle is considered.
(b)(color online) The theoretical molecule whose model is a
parallelogram satisfying the condition Eq.(\ref{relation}). The
upper and lower polygons are comprised of 2N particles so that their
total spins in the ground state is zero in the low temperature
respectively. The different colors represent the different coupling
constants.\label{V15}}
\end{figure}

To make up the theoretical molecular model, we need recall the model
of \{V15\} which can be considered as an isosceles triangle in a low
temperature first. The molecule \{V15\} is comprised of 15 $V^{4+}$
ions with each spin $s=1/2$, and the ions are arranged in a
quasispherical layered structure with a triangle sandwiched between
two hexagons as is shown in the Fig.\ref{V15}(a). Each hexagon of
\{V15\} consists of three pairs of strongly coupled spins with
$J_{1}\sim 800K$.\cite{i} Each spin of the $V^{4+}$ ions in the
central triangle is coupled with the spins in both hexagons with
$J_{2}=150K$ and $J_{3}=300K$, resulting in a very weak
exchange interaction between the spins within the central triangle
with $J_{0}=2.44K$.\cite{i,h} When the temperature is low enough, the
molecule \{V15\} has a much simpler approximation: the two hexagons
can be omitted because each total spin is zero, and the only survived part is just a simple Heisenberg spin triangle.\cite{g} It has been revealed that such a Heisenberg spin triangle model
is isosceles with the relationship
$J_{13}>J_{12}=J_{23}$ by measuring the local spin moments
configuration in an NMR experiment.\cite{d} So the Yangian symmetry
also exists in the \{V15\} molecule in a low temperature, and it can
be detected by measuring the local spin moments configuration.\\

Next we shall extend such a molecular model to the
four-spin system and give the prediction regarding the local spin moments
configuration. The assumed molecule is shown in the
Fig.\ref{V15}(b) whose structure is totally similar to the \{V15\}
molecule. The numbers of the ions in the upper and the lower
polygons are both even. And they are coupled to the special
parallelogram in the middle, which results in some special exchange
constants in the middle parallelogram satisfying the
Eq.(\ref{relation}). Similar to the \{V15\} molecule, such a
molecule can be viewed as a single parallelogram described in
Sec.\ref{sec-3}. We can calculate the local spin moments
configuration of this model in the NMR experiment. The local spin
moment is usually written as
$\mu_{i}=-g\mu_{B}\langle\phi_{g}|S^{3}_{i}|\phi_{g}\rangle$, where
the operator $S^{3}_{i}$ $(i=1,2,3,4)$ is the $z$-component of the spin
operator of the $i$-th particle, $g=2$, and $\mu_{B}$ is the Bohr
magneton. The state $|\phi_{g}\rangle$ is the ground state of the
system. Through direct calculation, we can obtain the local moments
of the four particles. It is highly nontrivial that the ground state
of a four-spin system is with $S=1$, so we mainly focus on such a
case and predict its measurement in the experiment. When the
exchange constants satisfy the relationship
$a_{12}>0,a_{13}<0,a_{13}<-2a_{12}$, the ground state is $\psi^3_{1,-1}$, and the corresponding local spin moments are
$\frac{9}{10}\mu_{B},\frac{1}{10}\mu_{B},\frac{1}{10}\mu_{B},\frac{9}{10}\mu_{B}$,respectively. So the observations are $\frac{1}{10}\mu_{B},\frac{9}{10}\mu_{B}$ when the NMR experiment is performed. In this way, we can tell whether there is the Yangian symmetry by measuring its local spin moments.

\section{Summary}
By investigating the experimental results on the molecular \{V6\},
we introduce a special form of the DM interaction operator, i.e. Yangian operator to describe the symmetry of the molecule. Meanwhile we extend
the symmetry operator to the four-spin system. We
find that the ground state may be no longer the singlet state in
a four-spin system determined by the Yangian symmetry. By choosing
suitable interaction constants, the state with spin $S=1$ can be the
ground state. When the ground state is with spin $S=1$, we extend
the DM interaction in the \{V6\} molecule to our new model ``\{V8\}''.
By investigating its magnetization, we find that it behaves similar
to the \{V6\} molecule. At last, we propose another theoretical molecular model and make a prediction regarding its local
spin moments configuration which might be measured by the NMR
experiment in principle.

\section*{Acknowledgments}
We thank M. G. Hu, Kai Niu and Ci Song for
helpful discussions. This work was supported by NSF of China (Grant
No. 10575053) and LuiHui Center for Applied Mathematics through the
joint project of Nankai and Tianjin Universities.

\appendix
\section{The mathematics of Yangian}\label{mathYang}
\setcounter{equation}{0}
\renewcommand{\theequation}{A\arabic{equation}}
Yangian algebras were established by
Drinfel'd.\cite{definition1,definition2,definition3} A Yangian is
formed by a set $\{\mathbf{I},\mathbf{Y}\}$ obeying the commutation
relations:
\begin{eqnarray}
&&[I_{\lambda},I_{\mu}]=c_{\lambda\mu\nu}I_{\nu},\qquad
[I_{\lambda},Y_{\mu}]=c_{\lambda\mu\nu}Y_{\nu},\label{definition1}\\
&&\lbrack[Y_{\lambda},[Y_{\mu},I_{\nu}]\rbrack-[I_{\lambda},[Y_{\mu},Y_{\nu}]]=\lambda
a_{\lambda\mu\nu\alpha\beta\gamma}\{I_{\alpha},I_{\beta},I_{\gamma}\},\label{definition2}\\
&&\lbrack[Y_{\lambda},Y_{\mu}],[I_{\sigma},Y_{\tau\tauup}]]+[[Y_{\sigma},Y_{\tauup}],[I_{\lambda},Y_{\mu}]]\nonumber\\
&&=\lambda(a_{\lambda\mu\nu\alpha\beta\gamma}c_{\sigma\tauup\nu}+a_{\sigma\tauup\nu\alpha\beta\gamma}c_{\lambda\mu\nu})\{I_{\alpha},I_{\beta},Y_{\gamma}\},\label{definition3}
\end{eqnarray}
where the set of $I_{\lambda}$ forms a simple Lie algebra
characterized by $c_{\lambda\mu\nu}$ and the repeated indices mean
summation. The definitions of $a_{\lambda\mu\nu\alpha\beta\gamma}$
and $\{x_{i},x_{j},x_{k}\}$ were given in the
Ref.\cite{definition1},
\begin{eqnarray}
a_{\lambda\mu\nu\alpha\beta\gamma}&=&\frac{1}{4!}c_{\lambda\alpha\sigma}c_{\mu\beta\tauup}c_{\nu\gamma\rho}c_{\sigma\tauup\rho},\\
\{x_{1},x_{2},x_{3}\}&=&\sum_{i\neq j\neq k}x_{i}x_{j}x_{k}.
\end{eqnarray}
In the present model
\begin{eqnarray}\label{Simplecase}
c_{\lambda\mu\nu}=i\epsilon_{\lambda\mu\nu}\quad (\lambda, \mu, \nu=1,2,3)
\end{eqnarray}
where $\epsilon_{\lambda\mu\nu}$ is the total anti-symmetric tensor. It turns out that Eq. (\ref{definition2}) is automatically satisfied. We then should simplify the Eq. (\ref{definition3}). Substituting Eq.(\ref{Simplecase}) into Eq.(\ref{definition3}) and considering all the possibilities of values of $\lambda, \mu, \nu, ...$ the Eq.(\ref{definition3}) reduces to
\begin{eqnarray}
[J_+,[J_3,J_+]]&=&\frac{\lambda}{4}I_+(J_+I_3-I_+J_3),\label{comminicate1}\\{}
[J_-,[J_3,J_-]]&=&\frac{\lambda}{4}I_-(J_-I_3-I_-J_3)\label{comminicate2}.
\end{eqnarray}
and
\begin{eqnarray}\label{Jacobi}
[J_\pm,[J_+,J_-]]+2[J_3,[J_3,J_\pm]]=\lambda \{2I_3(J_\pm I_3-I_\pm J_3)+I_\pm (I_\pm J_\mp-J_\pm I_\mp)\}
\end{eqnarray}
where $J_\pm=J_1\pm iJ_2, \quad I_\pm=I_1 \pm iI_2$.\\
It should be noted that
\begin{eqnarray}
J_\mp I\pm -I_\mp J_\pm =I_\pm J_\mp -J_\pm I_\mp.
\end{eqnarray}
We can prove that based on the Jacobian identities Eq. (\ref{Jacobi}) is satisfied. Therefore only Eq. (\ref{definition1}) and either (\ref{comminicate1}) or (\ref{comminicate2}) are independent. The similar relations for SU(3) and SO(5) occur. Their forms are complicated , so we shall not explain them here.
In the system comprised of $n$ particles, if the operator
$\mathbf{I}$ is taken as the total spin operator
$\mathbf{I}=\sum^{n}_{i=1}\mathbf{S}_{i}$, the Yangian operator
$\mathbf{Y}$ can be realized in terms of
\begin{eqnarray}\label{Yang}
\mathbf{Y}=\sum^{n}_{i=1}
u_i\mathbf{S}_{i}+i\sum^{n}_{i<j}(\mathbf{S}_{i}\times
\mathbf{S}_{j}),
\end{eqnarray}
where $u_i$ can be arbitrary parameters and $\mathbf{S}_{i}$ is
the spin of the $i$-th particle. It can be verified that the
operators $\mathbf{I}$ and $\mathbf{Y}$ satisfy the
Eq.(\ref{definition1}), Eq.(\ref{definition2}) and Eq.(\ref{definition3}), i.e. they form a Yangian. A set $\{Q=Y^2,I^2,I_3\}$ forms commuting set, so we use $Q$, $I^2$ and $I_3$ characterize a system. If $\lbrack[Q,H\rbrack]=0$, the Hamiltonian and $Q$ share the same eigenvalues. The Hamiltonian of Heisenberg model for $n$-spin system is usually written as:
\begin{eqnarray}\label{HM}
H=\sum^{n}_{i<j}a_{ij}\mathbf{S}_{i}\cdot\mathbf{S}_{j},
\end{eqnarray}
where $a_{ij}$ is the exchange constant between the $i$-th particle
and $j$-th particle. In contrast to the spin chain models we focus on few-body problem, say $n=3$ (\{V6\}) and $n=4$ (theoretical ``\{V8\}'' model). It is emphasized that the realization of Yangian shown by Eq. (\ref{Yang}) is not unique. There are other realization of Eq.(\ref{definition1}), Eq.(\ref{definition2}) and Eq.(\ref{definition3}), but Eq. (\ref{Yang}) is suitable to our discussion.

\section{Three-spin system with the Yangian symmetry}\label{thspin}
\setcounter{equation}{0}
\renewcommand{\theequation}{B\arabic{equation}}
For a three-spin system the Yangian operator $\mathbf{Y}$ is taken as Eq.(\ref{Yang}), and in the Hamiltonian Eq.(\ref{HM}) we have $n=3$. We need to calculate the eigenvalues and eigenstates of $Q$. The square of $\mathbf{Y}$ is given by
\begin{eqnarray}
Q&=&\sum^{3}_{i=1}u^2_iS^2_i+2(u_1u_2\mathbf{S_1}\cdot\mathbf{S_2}+u_2u_3\mathbf{S_2}\cdot\mathbf{S_3}+
u_1u_3\mathbf{S_1}\cdot\mathbf{S_3})\nonumber\\&&+2i(u_1-u_2+u_3)\mathbf{S_1}\cdot(\mathbf{S_2}\times\mathbf{S_3})-
\{S^2_1S^2_2+S^2_2S^2_3+S^2_1S^2_3\nonumber\\&&-(\mathbf{S_1}\cdot\mathbf{S_2}+\mathbf{S_2}\cdot\mathbf{S_3}+
\mathbf{S_1}\cdot\mathbf{S_3})+2S^2_1(\mathbf{S_2}\cdot\mathbf{S_3})+2S^2_3(\mathbf{S_1}\cdot\mathbf{S_2})\nonumber\\
&&-2S^2_2(\mathbf{S_1}\cdot\mathbf{S_3})-(\mathbf{S_1}\cdot\mathbf{S_2}+\mathbf{S_2}\cdot\mathbf{S_3}+
\mathbf{S_1}\cdot\mathbf{S_3})^2\nonumber\\&&+2[(\mathbf{S_1}\cdot\mathbf{S_2})(\mathbf{S_2}\cdot\mathbf{S_3})+
(\mathbf{S_2}\cdot\mathbf{S_3})(\mathbf{S_1}\cdot\mathbf{S_2})]\}\nonumber\\
\end{eqnarray}
And the usual Lie algebraic bases for three $\frac{1}{2}$-spins are as follows:
\begin{eqnarray}
|\phi_{\frac{3}{2},-\frac{3}{2}}\rangle&=&|\downarrow\downarrow\downarrow\rangle,\nonumber\\
|\phi_{\frac{3}{2},-\frac{1}{2}}\rangle&=&\frac{1}{\sqrt{3}}(|\uparrow\downarrow\downarrow\rangle+|\downarrow\uparrow\downarrow\rangle+
|\downarrow\downarrow\uparrow\rangle),\nonumber\\
|\phi_{\frac{3}{2},\frac{1}{2}}\rangle&=&\frac{1}{\sqrt{3}}(|\uparrow\uparrow\downarrow\rangle+|\uparrow\downarrow\uparrow\rangle+
|\downarrow\uparrow\uparrow\rangle),\nonumber\\
|\phi_{\frac{3}{2},\frac{3}{2}}\rangle&=&|\uparrow\uparrow\uparrow\rangle,\nonumber\\
|\phi^{\prime}_{\frac{1}{2},\frac{1}{2}}\rangle&=&\frac{1}{\sqrt{6}}(|\downarrow\uparrow\uparrow\rangle+|\uparrow\downarrow\uparrow\rangle-2
|\uparrow\uparrow\downarrow\rangle),\nonumber\\
|\phi^{\prime}_{\frac{1}{2},-\frac{1}{2}}\rangle&=&\frac{1}{\sqrt{6}}(|\uparrow\downarrow\downarrow\rangle+|\downarrow\uparrow\downarrow\rangle-2
|\downarrow\downarrow\uparrow\rangle),\nonumber\\
|\phi_{\frac{1}{2},\frac{1}{2}}\rangle&=&\frac{1}{\sqrt{2}}(|\downarrow\uparrow\uparrow\rangle+|\uparrow\downarrow\uparrow\rangle),\nonumber\\
|\phi_{\frac{1}{2},-\frac{1}{2}}\rangle&=&\frac{1}{\sqrt{2}}(|\uparrow\downarrow\downarrow\rangle+|\downarrow\uparrow\downarrow\rangle).\nonumber\\
\end{eqnarray}

It can be easily verified that:
\begin{eqnarray}
Q|\phi_{\frac{3}{2},m}\rangle&=&[\frac{3}{4}(u^{2}_{1}+u^{2}_{2}+u^{2}_{3})+\frac{1}{2}(u_{1}u_{2}+u_{2}u_{3}+u_{1}u_{3})-1]|\phi_{\frac{3}{2},m}\rangle,\nonumber\\
Q|\phi^{\prime}_{\frac{1}{2},m}\rangle&=&[\frac{3}{4}(u^{2}_{1}+u^{2}_{2}+u^{2}_{3})+\frac{1}{2}u_{1}u_{2}-u_{2}u_{3}-u_{1}u_{3}-\frac{7}{4}]|\phi^{\prime}_{\frac{1}{2},m}\rangle\nonumber\\
&&-\frac{\sqrt{3}}{2}(u_{1}-u_{2}+1)(u_{3}+1)|\phi_{\frac{1}{2},m}\rangle,\nonumber\\
Q|\phi_{\frac{1}{2},m}\rangle&=&-\frac{\sqrt{3}}{2}(u_{1}-u_{2}-1)(u_{3}-1)|\phi^{\prime}_{\frac{1}{2},m}\rangle
+[\frac{3}{4}(u_{1}-u_{2})^{2}+\frac{3}{4}u^{2}_{3}-\frac{3}{4}]|\phi^{\prime}_{\frac{1}{2},m}\rangle.\nonumber\\
\end{eqnarray}

We would like to note the difference between Lie algebra and Yangian in diagonalizing process. Suppose a Hamiltonian, in general, the eigenfunction of $H$ is written as $|\psi\rangle=\sum_Ic_I|\psi_I\rangle$ where $|\psi_I\rangle$ is the eigenfunction of $I^2$. However, if $[H,\mathbf{Y}]=0$, both $H$ and $\mathbf{Y}$ share the same wave function, because $\mathbf{Y}$ is an operator acting on tensor space which is much larger than Lie algebra space. For this reason, for symmetric $a_{ij}$ in Eq. (\ref{HM}) we choose symmetric $Q$ which leads to $u_2=u_1+u_3$. If $u_1=u_2=u_3=0$, the Yangian operator reduces to the special form fo the DM interaction operator Eq. (\ref{Yangian}). In this special case, we can easily obtain the eigenvalues and the eigenvectors of $Q$ as:
\begin{eqnarray}
Q|\phi_{\frac{3}{2},m}\rangle=-1|\phi_{\frac{3}{2},m}\rangle,\nonumber\\
Q|\psi^{\alpha}_{\frac{1}{2},m}\rangle=-\frac{1}{4}|\psi^{\alpha}_{\frac{1}{2},m}\rangle,\nonumber\\
Q|\psi^{beta}_{\frac{1}{2},m}\rangle=-\frac{9}{4}|\psi^{\beta}_{\frac{1}{2},m}\rangle,\nonumber\\
\end{eqnarray}

where $m$ is the magnetic quantum number, and the eigenvectors are

\begin{eqnarray}
|\psi^{\alpha}_{\frac{1}{2},m}\rangle&=&-\frac{1}{2}|\phi^{\prime}_{\frac{1}{2},m}\rangle+\frac{\sqrt{3}}{2}|\phi_{\frac{1}{2},m}\rangle,\nonumber\\
|\psi^{\beta}_{\frac{1}{2},m}\rangle&=&-\frac{\sqrt{3}}{2}|\phi^{\prime}_{\frac{1}{2},m}\rangle+\frac{1}{2}|\phi_{\frac{1}{2},m}\rangle.\nonumber\\
\end{eqnarray}

When $m=-1/2$ the two states with total spin $S=1/2$ are reduced to the
Eq.(\ref{states1}) and Eq.(\ref{states2}). We can see that the
operator $Q$ just mixes the states with the same numbers of
$S^2,S_{z}$. To let $Q$ and the Hamiltonian share the same
eigenstates  i.e.
\begin{eqnarray}
H|\phi_{\frac{3}{2},m}\rangle&=&E_{\frac{3}{2},m}|\phi_{\frac{3}{2},m}\rangle,\nonumber\\
H|\psi^{\alpha}_{\frac{1}{2},m}\rangle&=&E^{\alpha}_{\frac{1}{2},m}|\psi^{\alpha}_{\frac{1}{2},m}\rangle,\nonumber\\
H|\psi^{\beta}_{\frac{1}{2},m}\rangle&=&E^{\beta}_{\frac{1}{2},m}|\psi^{\beta}_{\frac{1}{2},m}\rangle\nonumber\\
\end{eqnarray}
by direct calculation, we get the relationship $a_{12}=a_{23}$
in the Hamiltonian(\ref{HM}) and the
corresponding energy levels $E_{\frac{1}{2},-\frac{1}{2}}^{\alpha}=\frac{1}{4}J_{13}-J_{12}$,
$E_{\frac{1}{2},-\frac{1}{2}}^{\beta}=-\frac{3}{4}J_{13}$,
$E_{\frac{3}{2},-\frac{3}{2}}=\frac{J_{12}}{2}+\frac{J_{13}}{4}$.\\

\section{The four-spin system with the Yangian
symmetry}\label{fspin}
\setcounter{equation}{0}
\renewcommand{\theequation}{C\arabic{equation}}
For a system comprised of four $\frac{1}{2}$-spins, the
similar process for $n=4$ will be performed. In the Yangian operator
Eq.(\ref{Yang}) and the Hamiltonian Eq.(\ref{HM}) we take
$n=4$. The operator $Q$ in them is
\begin{eqnarray}
Q&=&\sum^{4}_{i=1}u^2_iS^2_i+2\sum_{i,j(i<j)}^4u_iu_j\mathbf{S_i}\cdot\mathbf{S_j}\nonumber\\&&+2i[(u_1-u_2+u_3)\mathbf{S_1}\cdot(\mathbf{S_2}\times\mathbf{S_3})+
(u_1-u_2+u_4)\mathbf{S_1}\cdot(\mathbf{S_2}\times\mathbf{S_4})\nonumber\\&&+(u_1-u_3+u_4)\mathbf{S_1}\cdot(\mathbf{S_3}\times\mathbf{S_4})+
(u_2-u_3+u_4)\mathbf{S_2}\cdot(\mathbf{S_3}\times\mathbf{S_4})]\nonumber\\
&&-\{\sum_{i,j(i<j)}^4S^2_iS^2_j-[\sum_{i,j(i<j)}^4(\mathbf{S_{i}}\cdot\mathbf{S_{j}})]^2-\sum_{i,j(i<j)}^4(\mathbf{S_{i}}\cdot\mathbf{S_{j}})\nonumber\\
&&+2[S_1^2(\mathbf{S_2}\cdot\mathbf{S_3}+\mathbf{S_2}\cdot\mathbf{S_4}+\mathbf{S_3}\cdot\mathbf{S_4})+S^2_2(\mathbf{S_3}\cdot\mathbf{S_4}-\mathbf{S_1}\cdot\mathbf{S_3}\nonumber\\
&&-\mathbf{S_1}\cdot\mathbf{S_4})+S^2_3(\mathbf{S_1}\cdot\mathbf{S_2}-\mathbf{S_1}\cdot\mathbf{S_4}-\mathbf{S_2}\cdot\mathbf{S_4})\nonumber\\
&&+S^2_4(\mathbf{S_1}\cdot\mathbf{S_2}+\mathbf{S_1}\cdot\mathbf{S_3}+\mathbf{S_2}\cdot\mathbf{S_3})]\nonumber\\
&&+2[(\mathbf{S_2}\cdot\mathbf{S_3})(\mathbf{S_1}\cdot\mathbf{S_2})+(\mathbf{S_1}\cdot\mathbf{S_2})(\mathbf{S_2}\cdot\mathbf{S_3})]\nonumber\\
&&+2[(\mathbf{S_2}\cdot\mathbf{S_4})(\mathbf{S_1}\cdot\mathbf{S_2})+(\mathbf{S_1}\cdot\mathbf{S_2})(\mathbf{S_2}\cdot\mathbf{S_4})]\nonumber\\
&&+2[(\mathbf{S_3}\cdot\mathbf{S_4})(\mathbf{S_1}\cdot\mathbf{S_3})+(\mathbf{S_1}\cdot\mathbf{S_3})(\mathbf{S_3}\cdot\mathbf{S_4})]\nonumber\\
&&+2[(\mathbf{S_3}\cdot\mathbf{S_4})(\mathbf{S_2}\cdot\mathbf{S_3})+(\mathbf{S_2}\cdot\mathbf{S_3})(\mathbf{S_3}\cdot\mathbf{S_4})]\nonumber\\
&&+2[(\mathbf{S_1}\cdot\mathbf{S_3})(\mathbf{S_2}\cdot\mathbf{S_4})-(\mathbf{S_1}\cdot\mathbf{S_4})(\mathbf{S_2}\cdot\mathbf{S_3})+3(\mathbf{S_1}\cdot\mathbf{S_2})(\mathbf{S_3}\cdot\mathbf{S_4})]\}\nonumber\\
\end{eqnarray}
The usual Lie algebraic bases are standard:
\begin{eqnarray}
|\phi_{2,-2}\rangle&=&|\downarrow\downarrow\downarrow\rangle,\nonumber\\
|\phi_{2,-1}\rangle&=&\frac{1}{2}(|\uparrow\downarrow\downarrow\downarrow\rangle+|\downarrow\uparrow\downarrow\downarrow\rangle+
|\downarrow\downarrow\uparrow\downarrow\rangle+\downarrow\downarrow\downarrow\uparrow\rangle),\nonumber\\
|\phi_{2,0}\rangle&=&\frac{1}{\sqrt{6}}(|\uparrow\uparrow\downarrow\downarrow\rangle+\uparrow\downarrow\uparrow\downarrow\rangle+\uparrow\downarrow\downarrow\uparrow\rangle+\downarrow\uparrow\downarrow\uparrow\rangle
+\downarrow\downarrow\uparrow\uparrow\rangle+\downarrow\uparrow\uparrow\downarrow\rangle),\nonumber\\
|\phi_{2,1}\rangle&=&\frac{1}{2}(|\uparrow\uparrow\downarrow\uparrow\rangle+|\uparrow\downarrow\uparrow\uparrow\rangle+
|\downarrow\uparrow\uparrow\uparrow\rangle+\uparrow\uparrow\uparrow\downarrow\rangle),\nonumber\\
|\phi_{2,2}\rangle&=&|\uparrow\uparrow\uparrow\uparrow\rangle,\nonumber\\
|\phi^{1}_{1,1}\rangle&=&\frac{1}{2}(-|\uparrow\uparrow\uparrow\downarrow\rangle+|\uparrow\uparrow\downarrow\uparrow\rangle-
|\uparrow\downarrow\uparrow\uparrow\rangle-|\downarrow\uparrow\uparrow\uparrow\rangle),\nonumber\\
|\phi^{1}_{1,0}\rangle&=&\frac{1}{\sqrt{2}}(|\uparrow\uparrow\downarrow\downarrow\rangle-|\downarrow\downarrow\uparrow\uparrow\rangle),\nonumber\\
|\phi^{1}_{1,-1}\rangle&=&\frac{1}{2}(|\uparrow\downarrow\downarrow\downarrow\rangle+|\downarrow\uparrow\downarrow\downarrow\rangle-
|\downarrow\downarrow\uparrow\downarrow\rangle-|\downarrow\downarrow\downarrow\uparrow\rangle-),\nonumber\\
|\phi^{2}_{1,1}\rangle&=&\frac{1}{\sqrt{2}}(|\uparrow\downarrow\uparrow\uparrow\rangle-|\downarrow\uparrow\uparrow\uparrow\rangle),\nonumber\\
|\phi^{2}_{1,0}\rangle&=&\frac{1}{2}(|\uparrow\downarrow\uparrow\downarrow\rangle+|\uparrow\downarrow\downarrow\uparrow\rangle-
|\downarrow\uparrow\downarrow\uparrow\rangle-|\downarrow\uparrow\uparrow\downarrow\rangle),\nonumber\\
|\phi^{2}_{1,-1}\rangle&=&\frac{1}{\sqrt{2}}(|\uparrow\downarrow\downarrow\downarrow\rangle-|\downarrow\uparrow\downarrow\downarrow\rangle),\nonumber\\
|\phi^{3}_{1,1}\rangle&=&\frac{1}{\sqrt{2}}(|\uparrow\uparrow\uparrow\downarrow\rangle-|\uparrow\uparrow\downarrow\uparrow\rangle),\nonumber\\
|\phi^{3}_{1,0}\rangle&=&\frac{1}{2}(|\uparrow\downarrow\uparrow\downarrow\rangle+|\downarrow\uparrow\uparrow\downarrow\rangle-
|\downarrow\uparrow\downarrow\uparrow\rangle-|\uparrow\downarrow\downarrow\uparrow\rangle),\nonumber\\
|\phi^{3}_{1,-1}\rangle&=&\frac{1}{\sqrt{2}}(|\downarrow\downarrow\uparrow\downarrow\rangle-|\downarrow\downarrow\downarrow\uparrow\rangle),\nonumber\\
|\phi^{1}_{0,0}\rangle&=&\frac{1}{2\sqrt{3}}[2(|\uparrow\uparrow\downarrow\downarrow\rangle+|\downarrow\downarrow\uparrow\uparrow\rangle)-
(|\uparrow\downarrow\uparrow\downarrow\rangle\nonumber\\
&&+|\downarrow\uparrow\downarrow\uparrow\rangle+|\uparrow\downarrow\downarrow\uparrow\rangle+
|\downarrow\uparrow\uparrow\downarrow\rangle)],\nonumber\\
|\phi^{2}_{0,0}\rangle&=&\frac{1}{2}(|\uparrow\downarrow\uparrow\downarrow\rangle+|\downarrow\uparrow\downarrow\uparrow\rangle-
|\uparrow\downarrow\downarrow\uparrow\rangle-|\downarrow\uparrow\uparrow\downarrow\rangle).\nonumber\\
\end{eqnarray}

Then the action of $Q$ turns out that

\begin{eqnarray}
Q|\phi_{2,m}\rangle&=&[\frac{3}{8}(u_{1}+u_{2}+u_{3}+u_{4})^2+\frac{1}{4}(u_{1}-u_{2})^2-\frac{5}{2}\nonumber\\
&&+\frac{1}{8}(u_{1}+u_{2}-u_{3}-u_{4})^2+\frac{1}{4}(u_{3}-u_{4})^2]|\phi_{2,m}\rangle;\nonumber\\
Q|\phi^{1}_{1,m}\rangle&=&[\frac{1}{2}(u^{2}_{1}+u^{2}_{2}+u^{2}_{3}+u^{2}_{4})-\frac{9}{2}+\frac{1}{4}(u_{1}+u_{2}-u_{3}-u_{4})^2]|\phi^{1}_{1,m}\rangle\nonumber\\
&&-\frac{1}{\sqrt{2}}(u{1}-u_{2}+1)(u_{3}+u_{4}+2)|\phi^{2}_{1,m}\rangle+\frac{1}{\sqrt{2}}(u_{3}-u_{4}+1)(u_{1}-u_{2}-2)|\phi^{3}_{1,m}\rangle;\nonumber\\
Q|\phi^{2}_{1,m}\rangle&=&-\frac{1}{\sqrt{2}}(u{1}-u_{2}+1)(u_{3}+u_{4}+2)|\phi^{1}_{1,m}\rangle
+[\frac{1}{2}(u_{3}+u_{4})^2+\frac{1}{4}(u_{3}-u_{4})^2\nonumber\\
&&{}+\frac{3}{4}(u_{1}-u_{2})^2-1]|\phi^{2}_{1,m}\rangle
+\frac{1}{2}(u_{1}-u_{2}-1)(u_{3}-u_{4}+1)|\phi^{3}_{1,m}\rangle;\nonumber\\
Q|\phi^{3}_{1,m}\rangle&=&\frac{1}{\sqrt{2}}(u_{3}-u_{4}+1)(u_{1}-u_{2}-2)|\phi^{1}_{1,m}\rangle
+\frac{1}{2}(u_{1}-u_{2}-1)(u_{3}-u_{4}+1)|\phi^{2}_{1,m}\rangle\nonumber\\
&&+[\frac{1}{2}(u_{1}+u_{2})^2+\frac{1}{4}(u_{1}-u_{2})^2+\frac{3}{4}(u_{3}-u_{4})^2-1]|\phi^{3}_{1,m}\rangle;\nonumber\\
Q|\phi^{1}_{0,0}\rangle&=&\{\frac{1}{2}(u_{1}+u_{2}-u_{3}-u_{4}-2)(u_{1}+u_{2}-u_{3}-u_{4}+2)
+\frac{1}{4}[(u_{1}-u_{2})^2+(u_{3}-u_{4})^2-2]\}|\phi^{1}_{0,0}\rangle\nonumber\\
&&-\frac{\sqrt{3}}{2}(u_{1}-u_{2}+1)(u_{3}-u_{4}+1)|\phi^{2}_{0,0}\rangle;\nonumber\\
Q|\phi^{2}_{0,0}\rangle&=&-\frac{\sqrt{3}}{2}(u_{1}-u_{2}+1)(u_{3}-u_{4}+1)|\phi^{1}_{0,0}\rangle+\frac{3}{4}[(u_{1}-u_{2})^2+(u_{3}-u_{4})^2-2]|\phi^{2}_{0,0}\rangle;\nonumber\\
\end{eqnarray}
where $m$ is the magnetic quantum number. However, a symmetric $Q$ requires $u_{1}=u_{2}=u_{3}=u_{4}=0$. So the first term of the
Eq.(\ref{Yang}) must vanish, and the the Yangian operator is
identified with special form of the DM interaction operator. In terms of Eq. (C.3) the eigenvalues and the eigenstates of the $Q$ are:\\

\begin{eqnarray}\label{eigeneqation}
Q\left(\begin{array}{c}\phi_{2,m}\\\psi^{1}_{1,m}\\\psi^{2}_{1,m}\\\psi^{3}_{1,m}\\\psi^{+}_{00}\\\psi^{-}_{00}\end{array}\right)=\left(\begin{array}{cccccc}
-\frac{5}{2}&0&0&0&0&0\\0&-\frac{1}{2}&0&0&0&0\\0&0&-\frac{11}{2}&0&0&0\\0&0&0&-\frac{1}{2}&0&0\\
0&0&0&0&-1&0\\0&0&0&0&0&-3\end{array}\right)
\left(\begin{array}{c}\phi_{2,m}\\\psi^{1}_{1,m}\\\psi^{2}_{1,m}\\\psi^{3}_{1,m}\\\psi^{+}_{00}\\\psi^{-}_{00}\end{array}\right),
\end{eqnarray}

where the eigenstates are
\begin{eqnarray}
|\psi_{2,m}\rangle&=&|\phi_{2,m}\rangle,\nonumber\\
|\psi^1_{1,m}\rangle&=&\frac{2\sqrt{2}}{10}|\phi^1_{1,m}\rangle+\frac{1}{\sqrt{10}}|\phi^2_{1,m}\rangle+\frac{1}{\sqrt{10}}|\phi^3_{1,m}\rangle\nonumber\\
|\psi^2_{1,m}\rangle&=&\frac{1}{\sqrt{2}}|\phi^2_{1,m}\rangle-\frac{1}{\sqrt{2}}|\phi^3_{1,m}\rangle\nonumber\\
|\psi^3_{1,m}\rangle&=&-\frac{1}{\sqrt{5}}|\phi^1_{1,m}\rangle+\frac{2}{\sqrt{10}}|\phi^2_{1,m}\rangle+\frac{2}{\sqrt{10}}|\phi^3_{1,m}\rangle\nonumber\\
|\psi^+_{0,0}\rangle&=&\frac{\sqrt{3}}{2}|\phi^1_{0,0}\rangle+\frac{1}{2}|\phi^2_{0,0}\rangle,\nonumber\\
|\psi^-_{0,0}\rangle&=&-\frac{1}{2}|\phi^1_{0,0}\rangle+\frac{\sqrt{3}}{2}|\phi^2_{0,0}\rangle,\nonumber\\
\end{eqnarray}

In detail, the eigenstates with some definite magnetic quantum number $m$ is expanded as follows:

\begin{eqnarray}\label{eigenvec}
|\phi_{2,-2}\rangle&=&|\downarrow\downarrow\downarrow\downarrow\rangle,\nonumber\\
|\psi^{1}_{1,-1}\rangle&=&\frac{1}{2}\left(-|\uparrow\downarrow\downarrow\downarrow\rangle+|\downarrow\uparrow\downarrow\downarrow\rangle+
|\downarrow\downarrow\uparrow\downarrow\rangle-|\downarrow\downarrow\downarrow\uparrow\rangle\right),\nonumber\\
|\psi^{2}_{1,-1}\rangle&=&\frac{1}{\sqrt{20}}\left(3|\uparrow\downarrow\downarrow\downarrow\rangle+|\downarrow\uparrow\downarrow\downarrow\rangle-
|\downarrow\downarrow\uparrow\downarrow\rangle-3|\downarrow\downarrow\downarrow\uparrow\rangle\right),\nonumber\\
|\psi^{3}_{1,-1}\rangle&=&\frac{1}{\sqrt{20}}\left(|\uparrow\downarrow\downarrow\downarrow\rangle-3|\downarrow\uparrow\downarrow\downarrow\rangle+3
|\downarrow\downarrow\uparrow\downarrow\rangle-|\downarrow\downarrow\downarrow\uparrow\rangle\right),\nonumber\\
|\psi^{+}_{0,0}\rangle&=&\frac{1}{2\sqrt{3}}[\left(|\uparrow\uparrow\downarrow\downarrow\rangle+|\downarrow\downarrow\uparrow\uparrow\rangle\right)-2
(|\uparrow\downarrow\uparrow\downarrow\rangle,\nonumber\\
&&+|\downarrow\uparrow\downarrow\uparrow\rangle)+\left(|\uparrow\downarrow\downarrow\uparrow\rangle+
|\downarrow\uparrow\uparrow\downarrow\rangle\right)],\nonumber\\
|\psi^{-}_{0,0}\rangle&=&\frac{1}{2}[-\left(|\uparrow\uparrow\downarrow\downarrow\rangle+|\downarrow\downarrow\uparrow\uparrow\rangle\right)+\left(
|\uparrow\downarrow\downarrow\uparrow\rangle+|\downarrow\uparrow\uparrow\downarrow\rangle\right)].\nonumber\\
&&
\end{eqnarray}
From Eq. (\ref{eigeneqation}) it follows that for $S=1$ the two states for
$Q=-\frac{1}{2}$ are degenerate. Hence, we need another variable $\theta$ to distinguish from each other:

\begin{eqnarray}\label{comb}
|\psi^{1^{\prime}}_{1,-1}\rangle&=\cos\frac{\theta}{2}|\alpha^{1}_{1,-1}\rangle-\sin\frac{\theta}{2}|\alpha^{3}_{1,-1}\rangle\nonumber\\
|\psi^{3^{\prime}}_{1,-1}\rangle&=\sin\frac{\theta}{2}|\alpha^{1}_{1,-1}\rangle+\cos\frac{\theta}{2}|\alpha^{3}_{1,-1}\rangle\nonumber\\
\end{eqnarray}\\

As we have emphasized in Sec.\ref{sec-3}, we must take the six states $|\phi_{2,-2}\rangle,$
$|\psi^{1^{\prime}}_{1,-1}\rangle,$ $|\psi^{2}_{1,-1}\rangle,$
$|\psi^{3^{\prime}}_{1,-1}\rangle,$ $|\psi^{+}_{0,0}\rangle,$
$|\psi^{-}_{0,0}\rangle$ as the eigenstates of the Hamiltonian,
from which we can get the relationship for the interaction
constants:

\begin{eqnarray}\label{app-rela1}
a_{24}&=&\frac{1}{2}\left(a_{12}+2a_{13}-a_{34}\right)\nonumber,\\
a_{14}&=&\frac{1}{3}\left(a_{12}+2a_{13}\right)\nonumber,\\
a_{23}&=&\frac{1}{3}\left(2a_{12}-2a_{13}+3a_{34}\right),\nonumber\\
\end{eqnarray}

and the constants $a_{12}$, $a_{34}$ and $a_{13}$ satisfy the
equation:
\begin{eqnarray}\label{app-rela2}
\cos^{2}\frac{\theta}{2}\left(\frac{a_{12}}{2}-\frac{a_{34}}{2}\right)+\sin^{2}\frac{\theta}{2}\left(\frac{5}{2}a_{12}-\frac{5}{2}a_{34}\right)+\frac{1}{2}\sin\theta\left(-\frac{2}{3}a_{12}-2a_{34}+\frac{8}{3}a_{13}\right)
=0.\nonumber\\
\end{eqnarray}
Hence, the general four-spin Hamiltonian with the Yangian symmetry
can be written as:
\begin{eqnarray}\label{gen}
H&=&a_{12}\mathbf S_{1}\cdot\mathbf S_{2}+a_{34}\mathbf
S_{3}\cdot\mathbf S_{4}+a_{13}\mathbf S_{1}\cdot\mathbf
S_{3}\nonumber+\frac{1}{2}(a_{12}+2a_{13}-a_{34})\mathbf
S_{2}\cdot\mathbf S_{4} \\&&+\frac{1}{3}(a_{12}+2a_{13})\mathbf
S_{1}\cdot\mathbf
S_{4}\nonumber+\frac{1}{3}(2a_{12}-2a_{13}+3a_{34})\mathbf
S_{2}\cdot\mathbf S_{3},\\
\end{eqnarray}
where $a_{12}$, $a_{13}$ and $a_{34}$ satisfy the
Eq.(\ref{app-rela2}). If we take the viable $\theta$ to be zero,
then the Eq.(\ref{app-rela1})and(\ref{app-rela2}) reduced to the
Eq.(\ref{relation}) which leads to the special parallelogram model, and
the Hamiltonian Eq.(\ref{gen}) reduces to Eq.(\ref{4HM}).


\begin{thebibliography}{99}
\bibitem{demon1} In the usual materials of \{V6\}, the isosceles
triagle is characterized as $J_{12}=J_{13}$. In this paper, we make
a different note as $J_{12}=J_{23}$ for the convenience to introduce
the new operator.
\bibitem{k} Here we have assumed that the Landau-Zener
tuneling is adiabatic, so the state $|2,-1\rangle$ will directly
transit to the state $|2,1\rangle$ through the ``crossing point''
and the state $|2,0\rangle$ is not affected. Under this assumption,
it seems that there is no difference between \{V8\} and \{V6\} in
the magnetization.
\end{thebibliography}
\end{document}